\documentclass[longauth,letter]{aa}  
\usepackage{graphicx}
\usepackage{natbib}

\newcommand{\sub}[1]{_{\rm #1}}

\newcommand{\CII}{[C\,{\sc ii}]}

\newcommand{\HII}{H\,{\sc ii}}
\newcommand{\hoplus}{H$_2$O$^+$}
\newcommand{\kms}{km~s$^{-1}$}
\newcommand{\changed}{}
\newcommand{\thirdchanged}{}
\newcommand{\fourthchanged}{}

\usepackage{txfonts}
\begin{document}
\title{Detection of interstellar oxidaniumyl: 
abundant \hoplus{} towards the star-forming regions
DR21, Sgr~B2, and NGC6334\thanks{{\it Herschel} is an ESA space observatory with science instruments 
provided by European-led Principal Investigator consortia and with 
important participation from NASA.}}
\author{V.~Ossenkopf\inst{1,2}, H.S.P.~M\"uller\inst{1}, D.C.~Lis\inst{3},
P.~Schilke\inst{1,4}, T.A.~Bell\inst{3}, S.~Bruderer\inst{8}, E.~Bergin\inst{5},
C.~Ceccarelli\inst{6}, C.~Comito\inst{4}, J.~Stutzki\inst{1}
% ...
}

\institute{
I. Physikalisches Institut der Universit\"at 
zu K\"oln, Z\"ulpicher Stra\ss{}e 77, 50937 K\"oln, Germany
\and 
SRON Netherlands Institute for Space Research, P.O. Box 800, 9700 AV 
Groningen, Netherlands
\and 
California Institute of Technology,  Pasadena, CA 91125 USA
\and 
Max-Planck-Institut f\"ur Radioastronomie, Auf dem H\"ugel 69, 53121, Bonn, Germany
\and 
University of Michigan, Ann Arbor, MI 48197 USA
\and 
Laboratoire d'Astrophysique de Grenoble, UMR 5571-CNRS, Universit\'e Joseph Fourier, Grenoble, France
\and 
Universit\'{e} de Bordeaux, Laboratoire d’Astrophysique de Bordeaux, France; CNRS/INSU, UMR 5804, Floirac, France
\and 
Institute of Astronomy, ETH Z\"urich, 8093 Z\"urich, Switzerland
% ...
}

\authorrunning{V.~Ossenkopf, H.S.P.~M\"uller, D.~Lis, P.~Schilke et~al.}
\titlerunning{Detection of interstellar oxidaniumyl}

\abstract
{}
{
We identify a prominent absorption feature at 1115 GHz, detected in first 
HIFI spectra towards high-mass star-forming regions, and interpret its astrophysical origin. 
}
{
The characteristic hyperfine pattern of the \hoplus{}
ground-state rotational transition, and the lack of other known 
low-energy transitions in this frequency range, identifies 
the feature as \hoplus{} absorption against
the dust continuum background and 
allows us to derive the velocity profile of the absorbing gas.
By comparing this velocity profile with velocity profiles
of other tracers in the DR21 star-forming region, we constrain
the frequency of the transition and the conditions
for its formation. 
}
{
In DR21, the velocity distribution of \hoplus{} matches that of 
the \CII{} line at 158\ ${\mu}$m 
and of OH cm-wave absorption, both stemming from the hot and dense 
clump surfaces facing the \HII{}-region and dynamically 
affected by the blister outflow. 
Diffuse foreground gas dominates the absorption towards Sgr~B2. 
The integrated intensity of the absorption line allows us to derive
lower limits to the \hoplus{} column density of 
$7.2\times 10^{12}$~cm$^{-2}$ in NGC~6334, $2.3\times 10^{13}$~cm$^{-2}$ in DR21, and 
$1.1\times 10^{15}$~cm$^{-2}$ in Sgr~B2. 
}
{}


\maketitle

\section{Introduction}

Oxidaniumyl or oxoniumyl \citep[][]{IUPAC}, the reactive water cation,
\hoplus{}, plays a crucial role {\fourthchanged in the chemical network describing the formation of
oxygen-bearing molecules} in UV irradiated parts of molecular clouds
\citep{vDh_Black,Gerin}.
It was 
identified at optical wavelengths in the tails of comets in the 1970's 
\citep{Fehrenbach,H2O+_Kohoutek_1974a,H2O+_Kohoutek_1974b},
but its detection in the general interstellar medium has proven elusive.

We report a detection of the ground-state rotational transition of \hoplus{}
in some of the first spectra 
taken with the HIFI instrument \citep{HIFI} on board the {\it Herschel} Space 
Observatory \citep{Herschel} during the performance verification
campaign and early science observations.
Section 2 briefly introduces the properties of the sources where
\hoplus{} was detected.
Section 3 summarises the spectroscopic data of the molecule.
The observations and the line identification are described in Sect. 4 {\fourthchanged 
and in Sect. 5 we discuss} the physical properties of the 
\hoplus{} absorption layer.

\section{The sources}
\label{sect_dr21}

We observed three massive Galactic star-forming/\HII{} regions
with very different properties. The DR21 star-forming region
is embedded in a ridge of dense molecular material
that obscures it at optical wavelengths. The embedded cluster drives a
violent bipolar outflow and creates bright photon-dominated 
(or photo-dissociation) regions (PDRs), visible as
clumps of 8~$\mu$m PAH emission in {\it Spitzer} IRAC maps
\citep{Marston} and showing up {\fourthchanged in emission lines from tracers} of irradiated hot gas, 
such as HCO$^+$, high-$J$ CO, atomic and ionised carbon, and atomic oxygen
\citep{Lane,Jakob}.
The eastern, blue-shifted outflow expands in a blister-like fountain, 
while the western, red-shifted outflow is confined to a small cone.


The Sgr~B2(M) and (N) cores are the most massive star-formation sites in
our Galaxy. The line of sight, located in the plane of the Galaxy,
passes through many spiral arm clouds and the 
extended envelope of Sgr~B2 itself. 
The foreground clouds display a very rich molecular
and atomic spectrum \citep[][]{Polehampton2007},
although they often have very low densities and column densities, 
characteristic of diffuse or translucent clouds. The envelope of Sgr~B2 itself 
includes hot, low density layers at both the ambient cloud velocity of 
64~\kms{}, and at 0~\kms{} \citep[][]{Ceccarelli2002}.
Many species detected along this line of sight have not been found elsewhere
and the exact origin of the molecular 
features is often ambiguous because of the overlapping radial velocities
\citep[e.g.,][]{Comito2003}.

NGC6334 is a nearby molecular cloud complex containing several concentrations
of massive stars at various stages of evolution. The far-infrared source ``I''
contains an embedded cluster of NIR sources
\citep{Tapia}. {\fourthchanged Four compact mm continuum sources are located}
near the geometric centre of the cluster \citep{Hunter}. Although NGC6334I is
not known to exhibit strong absorption lines, its OH absorption profiles
\citep{Brooks} reveal two molecular clouds along this line of sight,
one with velocities between $-15$ and 2~km\,s$^{-1}$, and the other
near 6~km\,s$^{-1}$. 

\section{The \hoplus{} spectroscopy}
\label{sect_molecule}

H$_2$O$^+$ is a radical with a $^2B_1$ electronic ground state and bond
lengths and angle slightly larger than H$_2$O.
Quantum-chemical calculations 
\citep{H2O+_ai_1989} yield a ground-state dipole moment of {\changed 2.4~D}.
The $B_1$ symmetry of the ground electronic state leads to a reversal of the 
{\it ortho} and {\it para} levels relative to water.

\begin{figure}
\centering
\includegraphics[angle=270.0,width=\columnwidth]{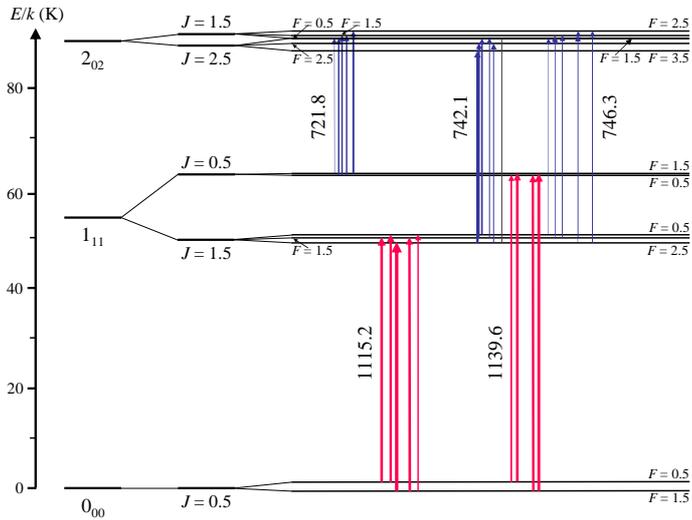}
\caption{
Energy level diagram of the lowest rotational levels of 
{\it ortho}-\hoplus{} and its radiative transitions.  
The fine structure transition frequencies are given in GHz.}
\label{energylevels}
\end{figure}

\begin{table}
\caption{Parameters of the hyperfine lines 
 $F' - F''$ in the observed 
$1_{11}-0_{00}, J=3/2-1/2$ ortho \hoplus{}
transition, including predicted frequencies, 
Einstein-$A$ and optical depth at low temperatures}
\label{line_parameter}
\begin{center}
\begin{tabular}{lrrrrrr}
\hline
$F' - F''$ & $\nu\sub{M\ddot{u}rtz}^a$ & $\nu\sub{Strahan}^a$ & $\nu\sub{OH-based}^b$ & $A$ &
$\int \tau dv$ \\
& $[$MHz$]$ & $[$MHz$]$ &  $[$MHz$]$ & $[$s$^{-1}]$ & $^c$ \\
\hline
$5/2-3/2$ &1115204.1 &1115175.8& 1115161& 0.031&	23.51 \\
$3/2-1/2$ &1115150.5 &1115122.0& 1115107& 0.017&	 8.67 \\
$3/2-3/2$ &1115263.2 &1115235.6& 1115221& 0.014&	 7.00 \\
$1/2-1/2$ &1115186.2 &1115158.0& 1115143& 0.027&	 6.96 \\
$1/2-3/2$ &1115298.9 &1115271.6& 1115257& 0.0035&	 0.88 \\
\hline
\end{tabular}
\end{center}
$^a$ {\thirdchanged Predictions based on \citet{H2O+_LMR_1986} and
\citet{H2O+_LMR_1998}. Nominal uncertainties are $\approx 2$~MHz 
but this is inconsistent with the discrepancy between the two 
predictions so that the actual uncertainty is unknown.}\\
$^b$ from the matching DR21 OH pattern by \cite{Guilloteau}\\
$^c$ $\int \tau\sub{low-\it T}/N\sub{H_2O^+} dv$ in $10^{-14}$ km s$^{-1}$ cm$^2$
\end{table}

The rotational spectrum was measured by 
laser magnetic resonance \citep{H2O+_LMR_1986,H2O+_LMR_1998}. 
Predictions of the $N_{K_aK_c} = 1_{11} - 0_{00}$, 
$J = 3/2 - 1/2$ fine structure component near 1115 GHz 
using the new parameters by \citet{H2O+_LMR_1998} are 
between 27.3 and 28.5~MHz higher than those calculated 
from \citet{H2O+_LMR_1986}, even though both articles claim 
to have reproduced the experimental data to $\sim$2~MHz.
The reanalysis of equivalent measurements of SH$^+$, by 
\citet{SH+_2009}, {\changed shows that this accuracy is
in principle achievable.} However, the large centrifugal distortion in H$_2$O$^+$ 
requires a large set of spectroscopic parameters to  
reproduce a comparatively small set of data; this may cause 
problems in the zero-field extrapolation. 
{\fourthchanged Moreover,} the frequencies of the two fine structure levels
of the $1_{11}$ rotational state in Table~V of \citet{H2O+_LMR_1998} 
agree precisely with those of the $F' = J'$, $F'' = J''$ hyperfine 
transitions. This can only be achieved when the calculated
frequencies are lower by 51.56 and 88.05~MHz, respectively,
since the respective hyperfine component is the lowest
in each case. Correcting the published frequencies
of the $J = 3/2 - 1/2$ fine structure component by 
51.56~MHz improves the agreement with \citet{H2O+_LMR_1986}.
The results are summarized in Table~\ref{line_parameter}.
Alternatively, we {\changed could} use the corrected frequencies of \citet{H2O+_LMR_1998} 
and arrive at values that are lower by about 23~MHz. 
This provides a rough estimate of the uncertainty in the predictions.
An H$_2$O$^+$ catalogue entry will be prepared for the CDMS \citep{CDMS2}
{\changed by carefully scrutinizing the available IR data summarised in 
\citet{H2O+_nu2_2008} with $\approx$~150~MHz uncertainties.} 

\section{Observations of the 1115~GHz ground-state transition}


\begin{table}
\caption{Summary of the observational parameters}
\label{tab_obs}
\begin{center}
\begin{tabular}{lrrr}
\hline
& DR21(C) & Sgr~B2(M) & NGC~6334 \\
\hline
RA (J2000) & 20h39m01.1s & 17h47m20.35s & 17h20m53.32s\\
DEC & 42$^\circ$19$'$43.0$''$ & -28$^\circ$23$'$03.0$''$ & -35$^\circ$46$'$58.5$''$\\
Mode & Load-chop$^2$ & DBS & DBS \\
$t\sub{int,source}$ & 150~s & 48~s & 48~s \\
$\sigma\sub{noise}^1$ & 0.07~K & 0.08~K & 0.08~K\\
\hline
\end{tabular}
\end{center}
$^1$ at native WBS resolution (1.1~MHz = 0.30~\kms{})\\
$^2$ OFF position=20h37m10s, 42$^\circ$37$'$00$''$
\end{table}

The \hoplus{} line was detected in DR21 during performance
verification observations {\fourthchanged of} the HIFI instrument, testing
spectral scans in the HIFI band 4b. Later science observations
of Sgr~B2 and NGC~6334 also confirmed the detection in these
sources using the identification and frequency assignment from
DR21. The main parameters
of the observations are summarised in Table~\ref{tab_obs}.
At 1115~GHz, the {\it Herschel} beam has 21$''$ HPBW. 

\begin{figure}
\includegraphics[angle=90.0,width=8.68cm]{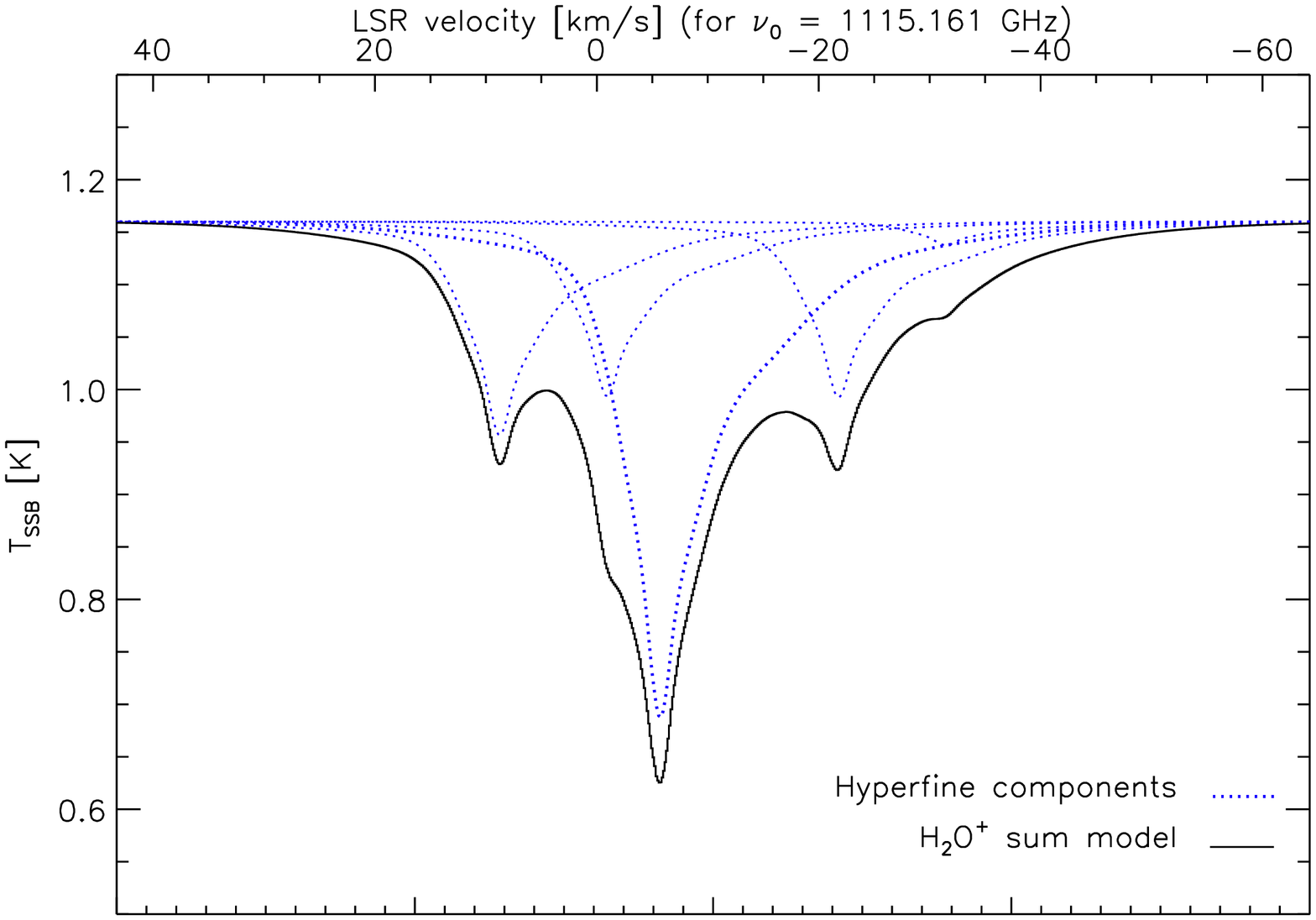}\\
\includegraphics[angle=90.0,width=\columnwidth]{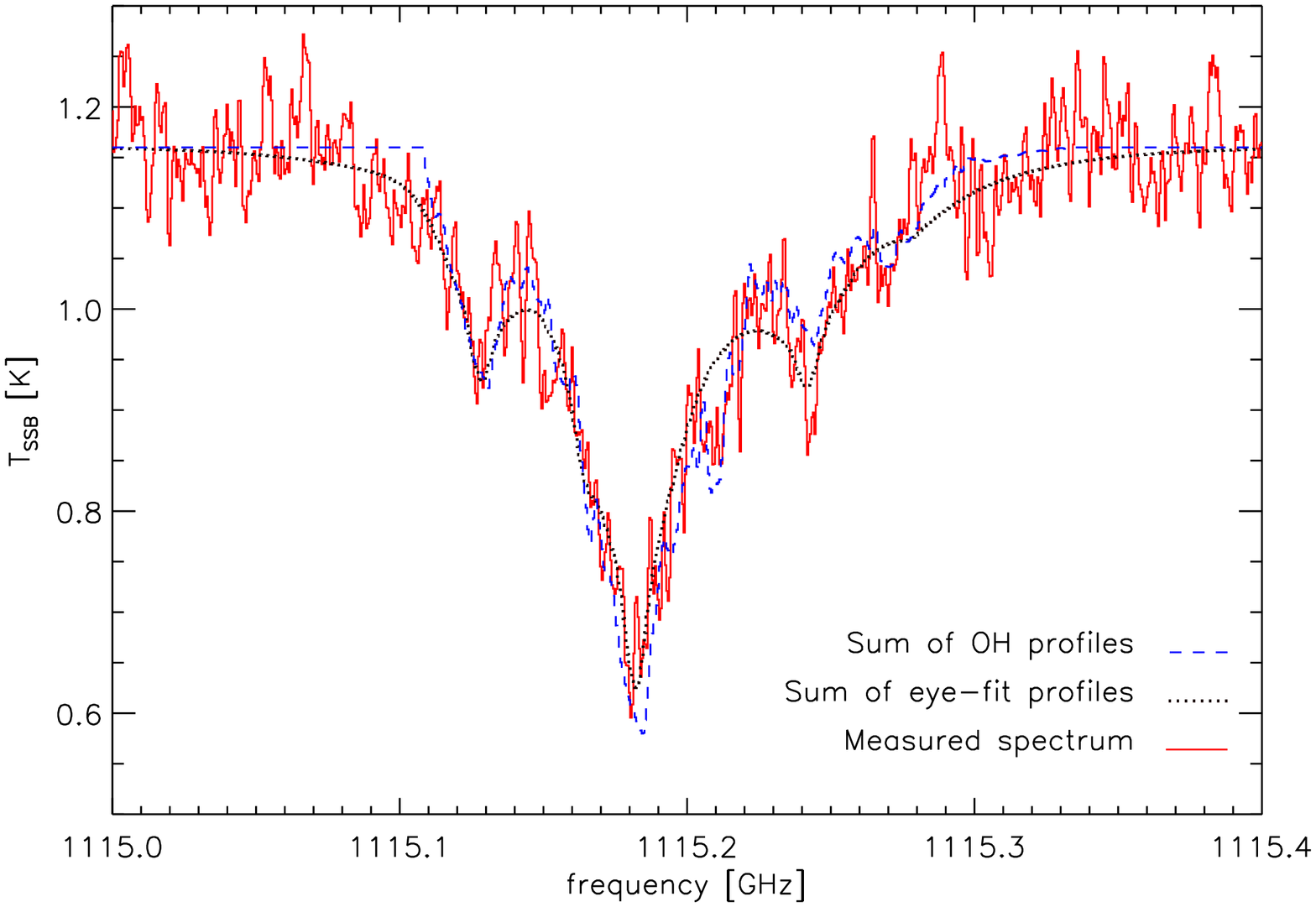}
\caption{Fit of the hyperfine multiplet of the 
\hoplus 1115~GHz line in DR21. The bottom panel shows the 
0.5~K absorption line superimposed on two different fit
profiles, one based on a 3-component Gaussian (see text) and 
the other one using the OH 6cm absorption spectrum
from \citet[][]{Guilloteau}. The top panel shows 
a breakdown of the fitted profile into its hyperfine 
constituents in the case of the 3-Gaussian profile.}
\label{fig_fit}
\end{figure}


The identification with \hoplus{} was straightforward in DR21 because
of the simple source velocity structure that cannot be confused with
the well resolved, characteristic hyperfine structure of the line. 
When fitting the line, one has to take into account that the 
line extinction begins to {\fourthchanged saturate,
with a maximum optical depth of 0.59 for DR21 and} 1.55 for Sgr~B2 (see below).
For DR21, we fitted the observed profile using an adjusted 
velocity profile with asymmetric wings. Because of the limited
signal-to-noise ratio, the fit was performed manually by adding
three Gaussian components of increasing width 
(see Fig. \ref{fig_fit}). 

\begin{figure}
\centering
\includegraphics[angle=90.0,width=\columnwidth]{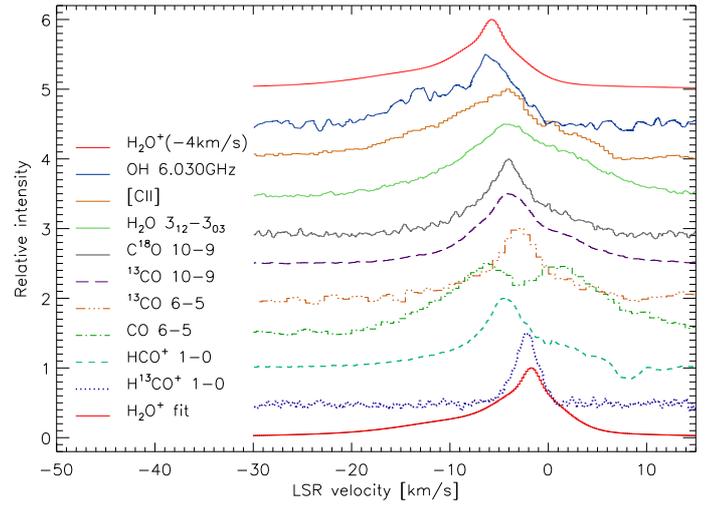}
\caption{Comparison of the fitted \hoplus velocity profile 
to other tracers observed in DR21 with similar beam size. The profiles
are normalised to a peak of unity and separated by multiples of 0.5 
from bottom to top. The fit (bottom line) used the \citet{H2O+_LMR_1986}
based line frequency prediction, the profile at the top is shifted by
-4.0~\kms{}, corresponding to a rest frequency lower by 15~MHz.}
\label{fig_mols}
\end{figure}

The resulting velocity distribution allows us to interpret the origin of
the absorbing material by comparing with the velocity distribution
of other species observed towards the same position with comparable 
beam size \citep[see][]{WADI1, Falgarone, vdT}. Figure \ref{fig_mols}
shows that the peak \hoplus{} velocity of $-1.7$~\kms{} is not seen in any
other tracer. The intrinsic velocity of the DR21 molecular ridge is
$-3.0$~km s$^{-1}$,
which is matched by the line centres of the H$^{13}$CO$^+$ 1--0, the CO
6--5, and the $^{13}$CO 6--5 transitions. The higher excitation lines
of $^{13}$CO, C$^{18}$O, H$_2$O, and the \CII{} line exhibit a slightly 
blue-shifted peak velocity of about $-5.0$~km s$^{-1}$. The \hoplus{}
profile exhibits a prominent, very broad blue wing. This is not present in
any of the molecular emission lines, but is found in the \CII{} profile and
the OH absorption spectrum measured by \citet{Guilloteau} towards
the same position.

\begin{figure}
\centering
\includegraphics[angle=90.0,width=\columnwidth]{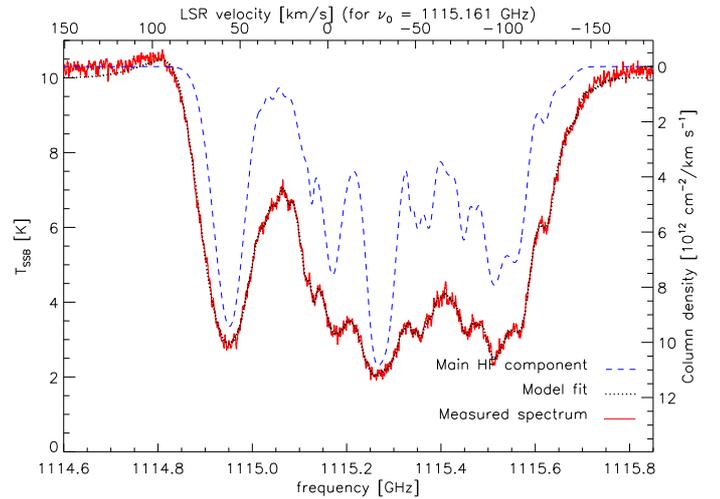}
\caption{Fit of the observed \hoplus line in Sgr~B2. The dashed line
visualises the velocity structure of the absorbers by plotting
the strongest hyperfine component on a linear column density
scale, i.e., without optical depth correction.}
\label{fig_sgrb2}
\end{figure}

\begin{figure}
\centering
\includegraphics[angle=90.0,width=\columnwidth]{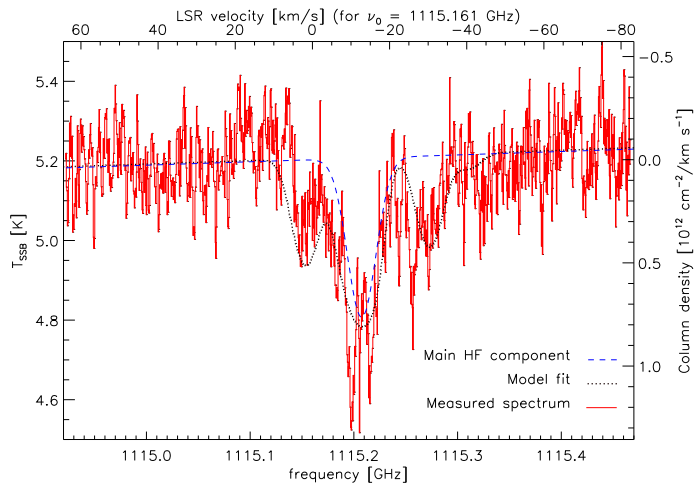}
\caption{Same as Fig.~\ref{fig_sgrb2}, but for NGC6334.}
\label{fig_ngc6334}
\end{figure}

To underline this good match, we have 
superimposed in Fig.~\ref{fig_fit} the absorption profile that would be
obtained by simply performing the hyperfine superposition of 
the 6.030~GHz OH absorption profile.
The match is as good as that achieved with the analytic profile
and even reproduces the small excursions at 1115.22 and 1115.27~GHz.
This indicates that OH and \hoplus{} occur in the same region and under the
same physical conditions.
The displacement of the fitted profile relative to the \CII{} and OH
profiles of about 4.0~km s$^{-1}$ is within the {\changed discrepancies 
between the different predictions} of the line frequency.
The astronomically determined line rest frequencies from
comparison with the OH line fall 15~MHz below the predicted
frequencies. As the line peak is very 
sharp, the accuracy of the frequency is probably better
than 2~MHz. Assuming a match with the \CII{} line instead,
would provide 
a larger uncertainty of $\approx$6~MHz.

The {\changed identification and the corrected frequencies} are then used to
analyse the line structures in Sgr~B2 and NGC~6334 (Figs.~\ref{fig_sgrb2}
and \ref{fig_ngc6334}).
In Sgr~B2, we see absorption at both the velocity
of its envelope and the velocities
of many foreground clouds, almost saturating the line. 
NGC6334 exhibits weak \hoplus{} absorption at 
$-13$~\kms{}.
This deviates from the OH absorption profile towards the 
source measured by \citet{Brooks}.
At velocities below $-10$~\kms{}, only some OH maser emission was found. This
might indicate that the observed \hoplus{} is not related to the foreground
material, but to hot gas in the direct vicinity of the continuum
sources. Alternatively, if we use the predicted frequencies from
{\thirdchanged \citet{H2O+_LMR_1986} in Table~\ref{line_parameter}}, the \hoplus{} absorption in NGC6334 is
centred on -9~\kms{}, in reasonable agreement with the OH absorption
at -8.2 km/s measured toward component F\footnote{A similar case is
reported by \citet{Gerin} for W31C. The source shows a complicated spectrum 
with multiple absorption components, but a {\fourthchanged closer} correlation with other
tracers is found when using the \citet{H2O+_LMR_1986} based frequencies.
{\thirdchanged {\fourthchanged A recent} detection of \hoplus in W3 IRS5 and AFGL2591
by Benz et al. (in prep.) seems to favour the frequency predictions by \citet{H2O+_LMR_1998}.}}.
{\changed At about $-9$~km\,s$^{-1}$, 
\citet{Beuther} also observed CH$_3$OH and NH$_3$ absorption 
towards the H\,{\sc ii} region.} 

\section{Discussion and outlook}


That \hoplus{} shows up in absorption against the 
dust continuum implies that the excitation of the molecule 
must be colder than the dust. 
As a reactive ion (see the discussion  by \citealt[][]{Black,Staeuber}
for CO$^+$), 
\hoplus{} is not expected to be in thermal equilibrium at the
kinetic temperature of the gas. Its excitation reflects either 
the chemical formation process or the radiative
coupling with the environment. 
From a single absorption line, {\fourthchanged one can only} provide
a lower limit to the \hoplus{} column density, assuming a low 
excitation temperature where basically all \hoplus{} resides in
the ground state, which is applicable to temperatures well below 
the upper level energy of 53~K. 

Table~\ref{line_parameter} provides the integral over the optical depth
of the hyperfine components in the low temperature limit.
For the overall $J=3/2-1/2$ fine structure transition, we 
obtain a line integrated optical depth of
$\int \tau d\nu/N\sub{H_2O^+}=4.70\,10^{-13}$ km s$^{-1}$ cm$^2$ 
per molecule, resulting in a lower limit to the 
\hoplus{} ground-state column densities 
of  $7.2\times 10^{12}$~cm$^{-2}$ for NGC~6334, $2.3\times 10^{13}$~cm$^{-2}$ 
for DR21, and $1.1\times 10^{15}$ cm$^{-2}$ for Sgr~B2.

These values are lower limits not only because of to the low-temperature
approximation, but also because they assume that the absorption occurs
in front of the continuum source and {\fourthchanged not within
the dusty cloud}, where the line absorption is partially 
compensated by dust emission. There may also be additional 
amounts of \hoplus{} in the para species that
would not contribute to the 1115~GHz line. Altogether, the total
\hoplus{} column density could be much higher than the lower limits
given here.

%


The excellent correlation between the \hoplus{} profile and the
OH absorption profile in DR21 indicates that both species occur
in the same thin layer of hot gas \citep{Jones} that directly
faces the \HII{} region at the blue-shifted blister outflow.
There is no obvious correlation with the distributions of CO, 
H$_2$O, or HCO$^+$. For Sgr~B2, we can clearly identify absorption
in multiple translucent foreground clouds. Their densities must be
high enough to produce
some molecular hydrogen, but low enough not to quickly destroy the
\hoplus{}.
For NGC6334, the gas component producing the \hoplus{} absorption
remains unidentified.

{\changed With the identification of H$_2$O$^+$ in the
interstellar medium, we provide a {\thirdchanged first step to quantifying}
an important intermediate node in the oxygen
chemical network, connecting OH$^+$ in diffuse clouds and at 
cloud boundaries, through H$_3$O$^+$, with water in denser
and cooler cloud parts.}
To obtain {\fourthchanged an estimate for the total \hoplus{} abundance,
we need to measure the excitation temperature of \hoplus{}.}
Observations of additional transitions
of \hoplus{}, such as those at 742~GHz, are therefore essential.

\begin{acknowledgements}
HIFI has been designed and built by a consortium of institutes and university departments from across
Europe, Canada and the United States under the leadership of SRON Netherlands Institute for Space
Research, Groningen, The Netherlands and with major contributions from Germany, France and the US.
Consortium members are: Canada: CSA, U.Waterloo; France: CESR, LAB, LERMA, IRAM; Germany:
KOSMA, MPIfR, MPS; Ireland, NUI Maynooth; Italy: ASI, IFSI-INAF, Osservatorio Astrofisico di Arcetri-
INAF; Netherlands: SRON, TUD; Poland: CAMK, CBK; Spain: Observatorio Astronómico Nacional (IGN),
Centro de Astrobiología (CSIC-INTA). Sweden: Chalmers University of Technology - MC2, RSS \& GARD;
Onsala Space Observatory; Swedish National Space Board, Stockholm University - Stockholm Observatory;
Switzerland: ETH Zurich, FHNW; USA: Caltech, JPL, NHSC.

      This work was supported by the German
      \emph{Deut\-sche For\-schungs\-ge\-mein\-schaft, DFG\/} project
      number Os~177/1--1.
HSPM is grateful to the Bundesministerium f\"ur Bildung und
Forschung (BMBF) for financial support aimed at maintaining the
Cologne Database for Molecular Spectroscopy, CDMS. This support has been
administered by the Deutsches Zentrum f\"ur Luft- und Raumfahrt (DLR).
D.C.L. is supported by the NSF,
award AST-0540882 to the Caltech Submillimeter Observatory.
A portion of this research was performed at the Jet Propulsion Laboratory, California Institute of Technology, under contract with the National Aeronautics and Space administration.
\end{acknowledgements}

\end{document}